\documentstyle[aps,epsf,floats]{revtex}            
\def\bq{\begin{equation}}
\def\eq{\end{equation}}
\def\ba{\begin{eqnarray}}
\def\ea{\end{eqnarray}}
\def\tev{\rm TeV}
\def\gev{\rm GeV}
\def\lsim{\mathrel{\raise.3ex\hbox{$<$\kern-.75em\lower1ex\hbox{$\sim$}}}}
\def\gsim{\mathrel{\raise.3ex\hbox{$>$\kern-.75em\lower1ex\hbox{$\sim$}}}}

\newcommand{ \slashchar }[1]{\setbox0=\hbox{$#1$}   
   \dimen0=\wd0                                     
   \setbox1=\hbox{/} \dimen1=\wd1                   
   \ifdim\dimen0>\dimen1                            
      \rlap{\hbox to \dimen0{\hfil/\hfil}}          
      #1                                            
   \else                                            
      \rlap{\hbox to \dimen1{\hfil$#1$\hfil}}       
      /                                             
   \fi}                                             %
\def\etmiss{\sla{E_T}}
\begin{document}
\thispagestyle{empty}
  
\newcommand{\sla}[1]{/\!\!\!#1}
\renewcommand{\small}{\normalsize} 

\preprint{
\font\fortssbx=cmssbx10 scaled \magstep2
\hbox to \hsize{
\hskip.5in \raise.1in\hbox{\fortssbx University of Wisconsin - Madison} 
\hfill\vtop{\hbox{\bf MADPH-99-1115}
            \hbox{hepph/9905423}
            \hbox{May 1999}} }
}
\title{\vspace{.5in}
Drell-Yan plus missing energy as a signal for extra dimensions
}

\author{T.~Han, D.~Rainwater and D.~Zeppenfeld}
\address{
Department of Physics, University of Wisconsin, Madison, WI 53706 
}
\maketitle
\begin{abstract}
We explore the search sensitivity for signals of large extra dimensions at 
hadron colliders via the Drell-Yan process $pp \to \ell^+\ell^- +\etmiss\ X$ 
($\ell = {\rm e},\mu$) where the missing transverse energy is the result of 
escaping Kaluza-Klein gravitons. We find that one is able to place exclusion 
limits on the gravity scale up to 560 GeV at the Fermilab Tevatron, and to 
4.0 (3.3) TeV at the CERN LHC, for $n=3$ (4) extra dimensions.
\end{abstract}


\section{Introduction}\label{sec:one}

The idea of the existence of extra spatial dimensions has fascinated physicists 
for nearly a century~\cite{kk}. A quantum theory of gravity seems to be 
consistent only via extension to extra dimensions~\cite{string}. If 
compactification of the extra dimensions occurs near the Grand Unification scale 
or the Planck scale $10^{16}-10^{18}\ \gev$, then the effects of quantum gravity 
would be accessible only at very high energy scales, beyond the reach of any 
collider experiment. In this case, one would have to understand the mechanism of 
weak scale stabilization (at about 100 GeV) against radiative corrections, the 
so-called hierarchy problem. Recently, a radical scenario has been 
advocated~\cite{add,bdn} wherein quantum gravity may become significant at a 
much lower energy scale ($M_S$), as low as ${\cal O}(\tev)$. The apparent large 
Planck scale ($M_{pl}\sim 10^{19}\ \gev$) is then attributed to the more rapid 
$1/r^{n+2}$ decrease of the gravitational force with distance in $n$ extra space 
dimensions. In terms of the large compactification size, $R\gg 1/M_S$, of the 
extra dimensions this leads to\footnote{The precise relation depends on
the convention for $R$ and $M_S$. It is taken to be 
$G_N^{-1}=4\pi R^n M_S^{n+2}$ in \cite{add}; and to be
$G_N^{-1}=8\pi R^n M_S^{n+2}$ in \cite{grw}
where $G_N^{-1}=M_{pl}^2$ is the Newton's constant.
For the convenience of calculations for physical quantities,
we take the normalization to be \cite{form}
$G_N^{-1}= R^n M_S^{n+2}/(4\pi)^{n/2}\Gamma(n/2)$.}
\bq
M_{pl}^2\sim R^n M_S^{n+2} \, .
\eq
Such a scenario alleviates the hierarchy problem by restating it as a 
geometrical one, namely understanding the size of the compact dimensions and the 
scale $M_S$ at which compactification occurs. This scenario could be realized in 
certain string formulations~\cite{joe}. Naturally, after compactification, there 
will be towers of Kaluza-Klein (KK) excitations with mass separation of 
${\cal O}(1/R)$. To avoid conflicts with the Standard Model (SM), it is assumed 
that the SM fields are stuck to a 4-dimensional hyper-surface, while only 
gravitons propagate in the extra dimensions. If we are interested in low-scale 
quantum gravity effects with $M_S\sim {\cal O}(\tev)$, the minimal scenario 
$n=1$ has been ruled out because $R\sim 10^8$~km, which would yield effects 
visible in planetary motion. For $n=2$, although there is no conflict with 
Newtonian gravity or astronomy for $R\sim $0.1 mm, the constraint from supernova 
cooling has put a bound on the scale up to $M_S>50\ \tev$~\cite{supernova}. One 
may even be able to push the scale up to $M_S>110\ \tev$ from the spectrum for 
diffuse gamma radiation~\cite{cosmo}. For $n>2$, there is no experimental or 
observational conflict with the theory.

Most interestingly, there will be significant consequences for low-energy 
phenomenology with this scenario. Although the coupling for an individual KK 
excitation is gravitationally suppressed, the cumulative effect from the tower 
of states is suppressed only by $1/M_S^{n+2}$. Generically, there are two 
classes of collider signals induced by the KK gravitons. First, real KK 
gravitons can be emitted off SM fields. Second, virtual KK gravitons may be 
exchanged between external SM fields. There have been many studies proposing 
searches for visible signatures of extra dimensions at 
colliders~\cite{grw,form,mpp,virtual,pheno,stringstates}. In this Letter, we 
study another process to explore the sensitivity to probe low-scale quantum 
gravity effects, Drell-Yan charged lepton pair production, which has previously 
been used as a powerful test of the Standard Model and has provided severe 
constraints on new physics. For the process of current interest, we look for the 
clean signal
\bq
pp \to \ell^+ \ell^- + \etmiss\  X,
\label{dymiss}
\eq
where the missing transverse energy $\etmiss$ is due to an escaping KK graviton. 
We calculate the signal rate and those of the leading SM backgrounds, and study 
the sensitivity to probe quantum gravity effects both at the Tevatron and at the 
LHC.


\section{Calculational Tools}\label{sec:two}

The signal subprocess is
\bq
q \bar{q} \to \ell^+ \ell^- + G_{KK}
\eq
where $\ell=e,\mu$, and the graviton $G_{KK}$ escapes the detector, resulting in 
missing energy. The signal can be described by fourteen tree level Feynman 
diagrams, seven each for $Z$ and $\gamma$ exchange, where a graviton is attached 
to each SM field and also to each SM vertex. Thus, off-shell effects of the $Z$ 
boson are fully included. This would be extremely tedious to calculate by hand, 
or even with the aid of a trace-based Feynman graph program, but we can make 
this straightforward by using the helicity amplitude method: summing the 
numerical values of individual Feynman graph amplitudes for a set of fixed 
external helicities and momenta, then squaring the summed amplitude and 
integrating over all possible helicities and phase space numerically. 

To do this, we constructed three new HELAS~\cite{HELAS} vertex routines, for the 
graviton-fermion-fermion, graviton-gauge boson-gauge boson and 
graviton-fermion-fermion-gauge boson vertices~\cite{form}. Coding of the matrix 
elements can then easily be done by hand. We have verified current conservation 
for the full matrix elements both analytically and numerically. The summation 
over KK states with different mass is carried out numerically based on the 
weight function in~\cite{form}.

As we are including photon interference effects, we must allow the $Z$ to be 
off-shell, and thus must include finite-width effects for the $Z$ propagators. 
This presents a problem for the calculation, as the graviton amplitudes are not 
gauge invariant for a $Z$ propagator including an imaginary piece. While we 
believe we can provide a formal prescription for properly including such a term, 
for the moment we must rely on an approximation that is known to be extremely 
reliable~\cite{width}: we do not include a finite width in the matrix elements, 
but instead multiply the summed-squared amplitude by an overall factor
\bq
{(\hat{s}-M_Z^2)^2 \over (\hat{s}-M_Z^2)^2 + (M_Z \Gamma_Z)^2}
\times
{(m_{\ell\ell}^2-M_Z^2)^2 \over (m_{\ell\ell}^2-M_Z^2)^2 + (M_Z \Gamma_Z)^2}
\eq
where $\hat{s}$ is the parton c.m.~energy squared and $m_{\ell\ell}$ the 
invariant mass of the lepton pair. This factor removes the zero-width 
propagators from the matrix elements and inserts Breit-Wigner resonances for the 
$Z$ boson. For phase space regions where the principal contribution comes from 
graphs where the incoming quark pair or outgoing lepton pair have invariant 
masses far from the $Z$ mass, this factor is essentially unity. When either pair 
is at or near the $Z$ mass, it approximates the full cross section by the 
correct resonant contributions.


\section{Results and Discussion}\label{sec:three}

For the numerical evaluations, we use CTEQ4L parton distribution 
functions~\cite{CTEQ4_pdf} and the EW parameters $m_Z = 91.19$~GeV, 
$m_t = 175.0$~GeV, $\sin^2 \theta_W = 0.2315$, and 
$G_F = 1.16639 \times 10^{-5}$~GeV$^{-2}$. We choose the factorization scale 
$\mu_f = E_T$ of the lepton system. We impose basic acceptance cuts for event 
identification, based on detector capability. These are
\ba
\label{eq:base1tev}
&&p_T(\ell)>15\ \gev , \quad |\eta_\ell| < 2.0 \quad {\rm for\ Tevatron},\\  
\label{eq:base1lhc}
&&p_T(\ell)>20\ \gev , \quad |\eta_\ell| < 2.5 \quad {\rm for\ LHC},
\ea
where $p_T(\ell)$ and $\eta_\ell$ are the transverse momentum and the 
pseudo-rapidity of a charged lepton.

A defining feature of the signal is the missing transverse momentum due to the 
escaping massive graviton. Also, most signal events are produced in association 
with a $Z$ boson. The irreducible SM background to this signal is from 
$\ell^+ \ell^- \bar\nu\nu$ events, which are dominated by $Z^{(*)}Z$ and 
$\gamma^*Z$ production, which we call ``Drell-Yan$+\nu\nu$''. We want to reduce 
the photon continuum from $\gamma^*Z$ events where the virtual photon produces 
a relatively low mass lepton pair, and therefore require  
\bq
\label{eq:base2}
\quad \sla{E_T} > 20\ \gev , \qquad m_{\ell\ell} > 10\ \gev \; .
\eq
For the Tevatron at $\sqrt{s} = 1.8$~TeV (Run I) with these cuts, this SM 
background is 9.4~fb; at $\sqrt{s} = 2.0$~TeV (Run II) it rises to 11~fb. The 
slightly higher $\sqrt{\hat{s}}$ and greatly increased luminosity expected for 
Run II allow us to impose an additional cut on the missing transverse energy, 
\bq
\label{eq:Tev2}
\quad \sla{E_T} > 100\ \gev \, ,
\eq
reducing the SM background to only 1.25~fb.

\begin{figure}[t]
\vspace*{0.5in}            
\begin{picture}(0,0)(0,0)
\includegraphics{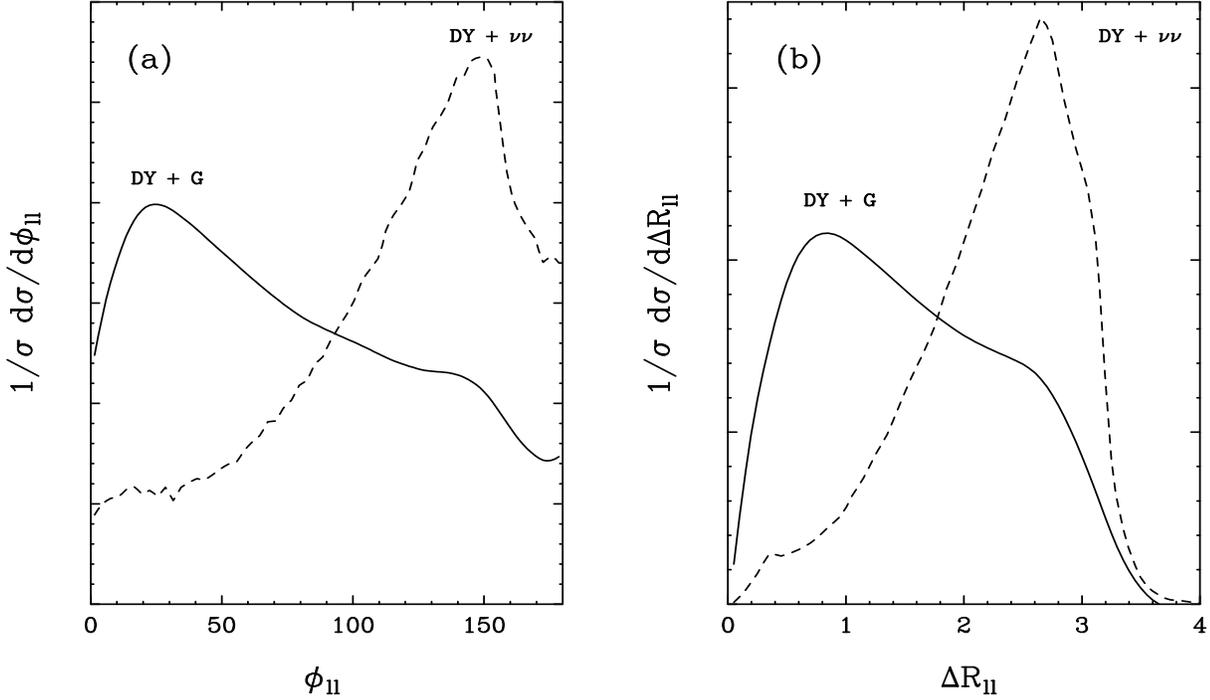}
\end{picture}
\vspace{8.5cm}
\caption{Normalized distributions of the lepton pair at the LHC: 
(a) azimuthal opening angle $\phi_{ll}$; (b) $\triangle R_{ll}$. 
The n = 3 graviton signal is shown with the solid curves, and the SM 
background with the dashed curves.}
\vspace*{0.2in}
\label{fig:ang_dist}
\end{figure}

\begin{figure}[t]
\vspace*{0.5in}            
\begin{picture}(0,0)(0,0)
\includegraphics{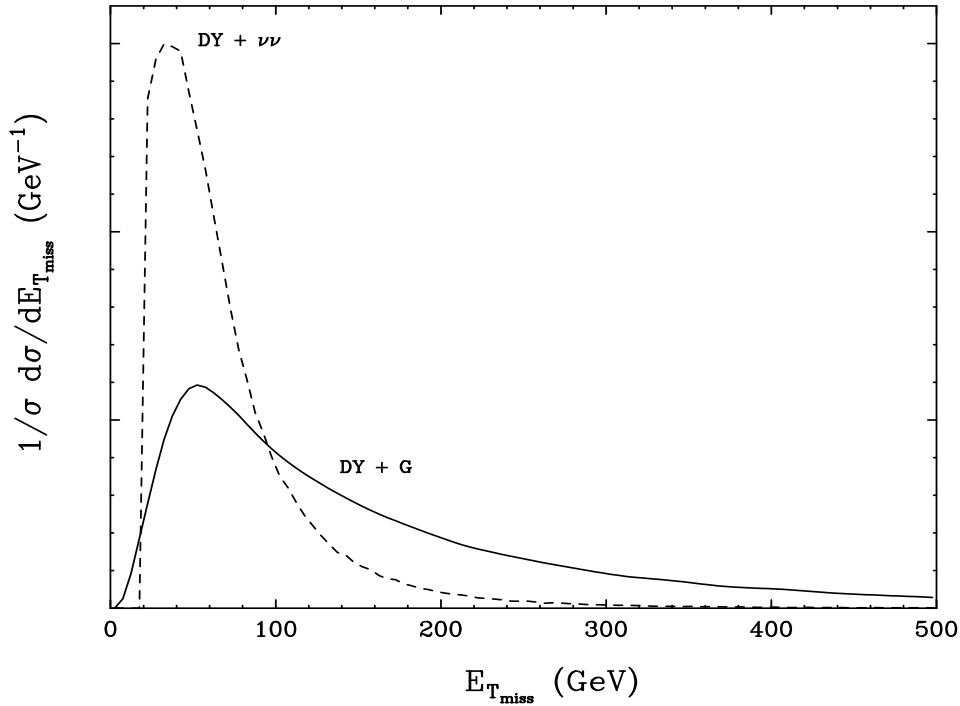}
\end{picture}
\vspace{8cm}
\caption{Normalized missing transverse momentum distributions of the $n = 3$ 
graviton signal (solid) and SM background (dashed) for the LHC.}
\label{fig:pT_miss}
\end{figure}

The much higher $\sqrt{\hat{s}}$ available at the LHC allows for the emission of 
much heavier gravitons and for a significant recoil of the dilepton system. This 
is reflected in the normalized dilepton angular distribution of 
Figure~\ref{fig:ang_dist}, for an $n=3$ signal and the DY$+\nu\nu$ background: 
(a) azimuthal opening angle $\phi_{\ell\ell}$; and (b) separation 
$\triangle R_{ll}=\sqrt{\phi_{\ell\ell}^2+(\eta_{\ell_1}-\eta_{\ell_2})^2}$. 
These plots show distinct differences for the signal and background. The final 
state leptons of the signal are preferentially emitted in the same direction, 
close to each other, while in the background the leptons tend to be more 
back-to-back. We can thus impose further cuts on the leptons, requiring
\bq
\label{eq:ang}
\phi_{\ell\ell} < 90^\circ \, , \quad \triangle{R_{\ell\ell}} < 1.2 \, .
\eq
This heavy graviton emission will also result in the signal exhibiting a much 
harder $\sla{E}_T$ spectrum, as shown in Fig.~\ref{fig:pT_miss}. We therefore 
suggest an even higher $\sla{E}_T$ cut for the LHC, 
\bq
\label{eq:LHC}
\sla{E}_T > 150\ \gev \, .
\eq
After the cuts of 
Eqs.~(\ref{eq:base1lhc},\ref{eq:base2},\ref{eq:ang},\ref{eq:LHC}), 
the SM background at the LHC is 2.8~fb.

A feature of any effective theory is that it is a low-energy approximation and 
can not be trusted at large energies, comparable to the defining scale of the 
theory. Applying this rule to the present case, our calculation requires some 
additional care for $n > 2$, as a non-trivial fraction of those signal events 
occur at center of mass energies $\sqrt{\hat{s}} > M_S$. 
A conservative estimate of the signal is obtained by discarding any events with 
$\sqrt{\hat{s}} > M_S$ \cite{grw}. 
More generally one could invoke form-factor damping of 
the high energy region. To obtain estimates of the string scales which can be 
probed experimentally we employ an iterative procedure, starting with a seed 
value for $M_S$ and throwing away all events with $\sqrt{\hat{s}} > M_S$. The 
resulting signal cross section is used to recompute a new exclusion limit for 
$M_S$, assuming a simple scaling behavior of the signal cross section, 
$\sigma_{signal}\sim 1/M_S^{n+2}$. Using the new exclusion limit as the upper 
bound on the allowable $\sqrt{\hat{s}}$ range, one obtains a stable exclusion 
limit in only 2-3 iterations. However, for $n > 4$ most of the signal events 
populate this unphysical region, and the overlap of the validity range of the 
effective theory with the range which would provide for a visible signal shrinks 
to zero. In this situation, one is unable to obtain any reliable estimate for a 
probe to $M_S$ via $\sla{E}_T$ signals. Instead, one would expect to first 
observe string excitations of particles states as a signature for new 
physics~\cite{stringstates}.

In Table~\ref{table1}, we summarize the sensitivity reach to the scale $M_S$ for 
$n=2-4$ via the process of Eq.~(\ref{dymiss}) at the Tevatron and the LHC. We 
see that this mode is a quite promising search channel. Although the Tevatron 
can impose some bounds, it is much more impressive to search for the signal at 
the LHC, its reach being several TeV for $n = 3,4$ extra dimensions. 

\begin{table} 
\begin{tabular}{lccc}
\phantom{generic} & n = 2 & n = 3 & n = 4 \\
\hline
Tevatron Run I @ 0.2 fb$^{-1}$  & 0.9 &      &     \\
Tevatron Run II @ 10 fb$^{-1}$  & 1.5 & 0.56 &     \\
LHC @ 100 fb$^{-1}$             & 5.3 & 4.0  & 3.3 \\
\end{tabular}
\vspace{0.15in}
\caption{String scale $M_S$ $95\%$ CL exclusion limits (TeV) at the Tevatron 
and LHC for $n = 2,3,4$ extra dimensions. For the Tevatron Run I value, we have 
imposed the cuts of Eqs.~(\ref{eq:base1tev},\ref{eq:base2}), and the additional 
cut of Eq.~(\ref{eq:Tev2}) for the Tevatron Run II, while for the LHC the cuts 
are given by Eqs.~(\ref{eq:base1lhc},\ref{eq:ang},\ref{eq:LHC}). The limits take 
into account the $\sqrt{\hat{s}}<M_S$ requirement discussed in the text.}
\label{table1}
\end{table}

A comparison with other studies at hadron colliders is in order. While the 
monojet$+\sla{E_T}$ signal~\cite{grw,mpp} from processes like 
$q\bar q \to g\ G_{KK}$ has the largest rate, it also has much more severe QCD 
backgrounds, mostly due to the mismeasurement of jets in the forward regions of 
the detector. Ref.~\cite{mpp} obtained results comparable to ours. For instance, 
at the LHC a $95\%$~CL limit is expected for $M_S=6.4\ (3.5)\ \tev$ with 
$n=2\ (4)$. On the other hand, Ref.~\cite{grw} reached a more impressive 
conclusion, claiming a 5$\sigma$ discovery for $M_S = 14$ (6.0)~TeV with $n = 2$ 
(4) extra dimensions\footnote{We have converted the scale $M_S$ to our 
normalization convention.}. Alternatively, the virtual $G_{KK}$ contribution to 
the DY process of $q\bar q \to \ell^+\ell^-$ can be also significant. It was 
found that a $95\%$~CL limit can be reached a scale $\Lambda\sim 1\ (6)\ \tev$ 
at the Tevatron and the LHC~\cite{virtual}, with a little dependence on $n$. 
However, one would have to introduce an additional assumption for a cutoff scale 
$\Lambda$above which the virtual KK tower is truncated, making a direct 
comparison of $M_S$ exclusion limits from external $G_{KK}$ production versus 
virtual exchange difficult.


\section{Conclusions}\label{sec:four}

We have explored the search sensitivity for signals of large extra dimensions at 
hadron colliders via the Drell-Yan  process $pp \to \ell^+\ell^- +\etmiss\ X$ 
($\ell = {\rm e},\mu$) where the missing transverse energy is the result of 
escaping Kaluza-Klein gravitons. This is a very clean channel for hadron 
collider physics. We find that one is able to place exclusion limits on the 
gravity scale up to 560 GeV at the Fermilab Tevatron, and up to 4.0 (3.3) TeV at 
the CERN LHC, for $n=3$ (4) extra dimensions. This reach is comparable to one 
found in previous studies of the monojet$+\sla{E_T}$ process~\cite{mpp} and for 
virtual contributions of $G_{KK}$ to the DY process~\cite{virtual}.


\acknowledgements
This research was supported in part by the University of Wisconsin 
Research Committee with funds granted by the Wisconsin Alumni Research 
Foundation and in part by the U.~S.~Department of Energy under 
Contract No.~DE-FG02-95ER40896.


\end{document}